\documentclass[conference,10pt]{IEEEtran}

\ifCLASSINFOpdf
 
\else
  
\fi

\usepackage{amsmath,amssymb,amsthm}
\usepackage{color}
\usepackage[noadjust]{cite}
\usepackage{graphicx}
\usepackage{epsfig}
\usepackage{breqn}
\usepackage[caption=false,font=footnotesize]{subfig}
\usepackage{array}

\usepackage{mathtools}
\makeatletter
\newcommand{\vast}{\bBigg@{4}}
\newcommand{\Vast}{\bBigg@{5}}

\makeatother

\newtheorem{theorem}{Theorem}[section]

\newtheorem{prop}[theorem]{Proposition}

\newtheorem{conjecture}[theorem]{Conjecture}

\allowdisplaybreaks
\newcommand\given[1][]{\:#1\vert\:}

\usepackage[margin=0.75in]{geometry}

\begin{document}
\bstctlcite{IEEEexample:BSTcontrol}

\title{\vspace{0.25in} Node Isolation of Secure Wireless Sensor Networks under a Heterogeneous Channel Model}



\author{\IEEEauthorblockN{Rashad Eletreby and Osman Ya\u{g}an}
\IEEEauthorblockA{Department
of Electrical and Computer Engineering and CyLab, \\
Carnegie Mellon University, Pittsburgh,
PA, 15213 USA\\
reletreby@cmu.edu, oyagan@ece.cmu.edu}}

%

\maketitle

\begin{abstract}
We investigate the secure connectivity of wireless sensor networks under a heterogeneous random key predistribution scheme and a heterogeneous channel model. In particular, we study a random graph formed by the intersection of an inhomogeneous random key graph with an inhomogeneous Erd\H{o}s-R\'enyi graph. The former graph is naturally induced by the heterogeneous random key predistribution scheme while the latter graph constitutes a heterogeneous on/off channel model; wherein, the wireless channel between a class-$i$ node and a class-$j$ node is on with probability $\alpha_{ij}$ independently. We present conditions (in the form of zero-one laws) on how to scale the parameters of the intersection model so that it has no isolated node with high probability as the number of nodes gets large. We also present numerical results to support these zero-one laws in the finite-node regime.
\end{abstract}


\IEEEpeerreviewmaketitle

\section{Introduction}
Wireless sensor networks (WSNs) consist of low-cost, low-power, small sensor nodes that are typically deployed randomly in large numbers enabling numerous applications such as military applications, health monitoring, environmental monitoring, etc \cite{Akyildiz_2002}. WSNs are typically deployed in hostile environments (e.g., battlefields), making it crucial to use cryptographic protection to secure sensor communications. In \cite[Chapter~13 and references therein]{Raghavendra_2004} and \cite{camtepe_2005}, authors review several key distribution schemes for WSNs and investigate their applicability given the classical constraints of a sensor node, namely: limited computational capabilities, limited transmission power, lack of a priori knowledge of deployment configuration, and vulnerability to node capture attacks. We refer the reader to \cite{Perrig_2004,Xiaojiang_2008} for a detailed analysis of security challenges in WSNs.

In \cite{Yagan/Inhomogeneous}, Ya\u{g}an introduced a new variation of the Eschenauer and Gligor (EG) key predistribution scheme \cite{Gligor_2002}, referred to as the heterogeneous key predistribution scheme. The heterogeneous key predistribution scheme accounts for the cases when the network comprises sensor nodes with varying level of resources and/or connectivity requirements, e.g., regular nodes vs. cluster heads, which is likely to be the case for many WSN applications \cite{Yarvis_2005}. According to this scheme, each sensor node belongs to a specific priority class and is given a number of keys corresponding to its class. More specifically, Given $r$ classes, a sensor node is classified as a class-$i$ node with probability $\mu_i$, resulting in a probability distribution $\pmb{\mu}=\{\mu_1,\mu_2,\ldots,\mu_r \}$ with $\mu_i >0, \text{ for } i=1,\ldots,r$ and $\sum_{i=1}^r \mu_i=1$. Sensors belonging to class-$i$ are each given $K_i$ keys selected uniformly at random (without replacement) from a key pool of size $P$. As with the EG scheme, pairs of sensors that share key(s) can communicate securely over an available channel after deployment.

Let $\mathbb{K}(n,\pmb{\mu},\pmb{K},P)$ denote the random graph induced by the heterogeneous key predistribution scheme described above, where $\pmb{K}=\{K_1,K_2,\ldots,K_r \}$ and $n$ denotes the number of nodes. A pair of nodes are adjacent as long as they share a key. 
This model is referred to as the {\em inhomogeneous} random key graph in \cite{Yagan/Inhomogeneous}; wherein, zero-one laws for the properties that $\mathbb{K}(n,\pmb{\mu},\pmb{K},P)$ i) has no isolated nodes and ii) is connected are established under the assumption of {\em full visibility}. Namely, it was assumed that all wireless channels are reliable and secure communications among participating nodes require only the existence of a shared key.

Our paper is motivated by the fact that the full visibility assumption is too optimistic and is not likely to hold in most WSN applications; e.g., the wireless medium of communication is often unreliable and sensors typically have limited communication ranges.
To that end, we study the secure connectivity of heterogeneous WSNs under a heterogeneous on/off communication model; wherein, the communication channel between two nodes of class-$i$ and class-$j$ is on with probability $\alpha_{ij}$. The heterogeneous on/off communication model induces the inhomogeneous Erd\H{o}s-R\'enyi (ER) graph \cite{devroye2014connectivity,BollobasJansonRiordan}, denoted hereafter by $\mathbb{G}(n,\pmb{\mu},\pmb{\alpha})$. The overall WSN can then be modeled by a random graph model formed by the intersection of an inhomogeneous random key graph and an inhomogeneous ER graph.
We denote the intersection graph $\mathbb{K}(n;\pmb{\mu},\pmb{K},P) \cap \mathbb{G}(n;\pmb{\mu},\pmb{\alpha})$ by $\mathbb{H}(n;\pmb{\mu},\pmb{K},P,\pmb{\alpha})$. 

Our main contribution is as follows.
We present conditions (in the form of zero-one laws) on how to scale the parameters of the intersection model $\mathbb{H}(n;\pmb{\mu},\pmb{K},P,\pmb{\alpha})$ so that it has no secure node which is isolated with high probability when the number of nodes $n$ gets large.
Our result generalizes several results in the literature, including the zero-one laws for absence of isolated nodes in inhomogeneous random key graphs intersecting homogeneous ER graphs \cite{Rashad/Inhomo}, and in homogeneous random key graphs intersecting homogeneous ER graphs \cite{Yagan/EG_intersecting_ER}.


We close with a word on notation and conventions in use. All limiting statements, including asymptotic equivalence are considered with the number of sensor nodes $n$ going to infinity. The random variables (rvs) under consideration are all defined on the same probability triple $(\Omega,\mathcal{F},\mathbb{P})$. Probabilistic statements are made with respect to this probability measure $\mathbb{P}$, and we denote the corresponding expectation by $\mathbb{E}$. The indicator function of an event $E$ is denoted by $\pmb{1}[E]$. We say that an event holds with high probability (whp) if it holds with probability $1$ as $n \rightarrow \infty$. For any discrete set $S$, we write $|S|$ for its cardinality.
In comparing
the asymptotic behaviors of the sequences $\{a_n\},\{b_n\}$,
we use
$a_n = o(b_n)$,  $a_n=\omega(b_n)$, $a_n = O(b_n)$, $a_n = \Omega(b_n)$, and
$a_n = \Theta(b_n)$, with their meaning in
the standard Landau notation. We also use $a_n \sim b_n$ to denote the asymptotic equivalence $\lim_{n \to \infty} {a_n}/{b_n}=1$.

\section{The Model}
We consider a network consisting of $n$ sensors labeled as $v_1, v_2, \ldots,v_n$. Each sensor node is classified into one of the $r$ classes, e.g., priority levels, according to a probability distribution $\pmb{\mu}=\{\mu_1,\mu_2,\ldots,\mu_r\}$ with $\mu_i >0$ for $i=1,\ldots,r$ and $\sum_{i=1}^r \mu_i=1$. Then, a class-$i$ node is assigned $K_i$ cryptographic keys selected uniformly at random and \textit{without replacement} from a key pool of size $P$. It follows that the key ring $\Sigma_x$ of node $x$ is an $\mathcal{P}_{K_{t_x}}$-valued random variable (rv) where $\mathcal{P}_{K_{t_x}}$ denotes the collection of all subsets of $\{1,\ldots,P\}$ with exactly $K_{t_x}$ elements and $t_x$ denotes the class of node $v_x$. The rvs $\Sigma_1, \Sigma_2, \ldots, \Sigma_n$ are then i.i.d. with
\begin{equation} \nonumber
\mathbb{P}[\Sigma_x=S \mid t_x=i]= \binom P{K_i}^{-1}, \quad S \in \mathcal{P}_{K_i}.
\nonumber
\end{equation}
Let $\pmb{K}=\{K_1,K_2,\ldots,K_r\}$ and assume without loss of generality that $K_1 \leq K_2 \leq \ldots \leq K_r$. Consider a random graph $\mathbb{K}$ induced on the vertex set $\mathcal{V}=\{v_1,\ldots,v_n\}$ such that a pair of distinct nodes $v_x$ and $v_y$ are adjacent in $\mathbb{K}$, denoted by $v_x \sim_K v_y$, if they have at least one cryptographic key in common, i.e.,
\begin{equation}
v_x \sim_K v_y \quad \text{if} \quad \Sigma_x \cap \Sigma_y \neq \emptyset.
\label{adjacency_condition}
\end{equation}

The adjacency condition (\ref{adjacency_condition}) defines the inhomogeneous random key graph denoted by $\mathbb{K}(n;\pmb{\mu},\pmb{K},P)$  \cite{Yagan/Inhomogeneous}. This model is also known in the literature  as
the {\em general random intersection graph}; e.g., see \cite{Zhao_2014,Rybarczyk,Godehardt_2003}. 
 The probability $p_{ij}$ that a class-$i$ node and a class-$j$ node are adjacent is given by
\begin{equation}
p_{ij}=1-\frac{\binom {P-K_i}{K_j}}{\binom {P}{K_j}}
\label{eq:osy_edge_prob_type_ij}
\end{equation}
as long as $K_i + K_j \leq P$; otherwise if $K_i +K_j > P$, we  have $p_{ij}=1$.
Let $\lambda_i$ denote the \textit{mean} probability that a class-$i$ node is connected to another node in $\mathbb{K}(n;\pmb{\mu},\pmb{K},P)$. We have
\begin{align}
\lambda_i & =\sum_{j=1}^r \mu_j p_{ij}.
 \label{eq:osy_mean_edge_prob_in_RKG}
\end{align}

We aim to investigate the performance of the heterogeneous key predistribution scheme without the \textit{full visibility} assumption \cite{Yagan/Inhomogeneous}. More precisely, to account for the possibility that communication channels between two nodes may not be available, e.g., due to deep fading, interference, etc., we assume a heterogeneous on/off channel model; wherein, the communication channel between two nodes of type-$i$ and type-$j$ is on with probability $\alpha_{ij}$. Consider a random graph $\mathbb{G}$ induced on the vertex set $\mathcal{V}=\{v_1,\ldots,v_n\}$ such that each node is classified into one of the $r$ classes, e.g., priority levels, according to a probability distribution $\pmb{\mu}=\{\mu_1,\mu_2,\ldots,\mu_r\}$ with $\mu_i >0$ for $i=1,\ldots,r$ and $\sum_{i=1}^r \mu_i=1$. Then, a distinct class-$i$ node $v_x$ and a distinct class-$j$ node $v_y$ are adjacent in $\mathbb{G}$, denoted by $v_x \sim_G v_y$, if $B_{xy}(\alpha_{ij})=1$ where $B_{xy}(\alpha_{ij})$ denotes a Bernoulli rv with success probability $\alpha_{ij}$. This adjacency conditions induces the inhomogeneous ER graph $\mathbb{G}(n;\pmb{\mu},\pmb{\alpha})$ on the vertex set $\mathcal{V}$, which has received some interest recently \cite{devroye2014connectivity,BollobasJansonRiordan}, and 
 would account for the fact that different nodes could have different radio capabilities, or could be deployed in locations with different channel characteristics. Although the on/off channel model may be considered too simple, it allows a comprehensive analysis of the properties of interest and is often a good approximation of more realistic channel models, e.g., the disk model \cite{Gupta99}. In fact, the simulations results in \cite{Yagan/EG_intersecting_ER} suggest that the connectivity behavior of the EG scheme under the on/off channel model is asymptotically equivalent to that under the disk model.


Our system model is obtained by the intersection of the inhomogeneous random key graph $\mathbb{K}(n;\pmb{\mu},\pmb{K},P)$ with the inhomogeneous ER graph $\mathbb{G}(n;\pmb{\mu},\pmb{\alpha})$. We denote the intersection graph by $\mathbb{H}(n;\pmb{\mu},\pmb{K},P,\pmb{\alpha})$, i.e., $\mathbb{H}(n;\pmb{\mu},\pmb{K},P,\pmb{\alpha}):=\mathbb{K}(n;\pmb{\mu},\pmb{K},P) \cap \mathbb{G}(n;\pmb{\mu},\pmb{\alpha})$. A distinct class-$i$ node $v_x$ is adjacent to a distinct class-$j$ node $v_y$ in $\mathbb{H}$ if and only if they are adjacent in both $\mathbb{K}$ \textit{and} $\mathbb{G}$. In words, the edges in $\mathbb{H}(n;\pmb{\mu},\pmb{K},P,\pmb{\alpha})$ represent pairs of sensors that share cryptographic key(s) {\em and} have a communication channel in between that is on, and hence can communicate securely. Therefore, studying the connectivity properties of $\mathbb{H}(n;\pmb{\mu},\pmb{K},P,\pmb{\alpha})$ amounts to studying the secure connectivity of heterogeneous WSNs under the heterogeneous on/off channel model.

To simplify the notation, we let $\pmb{\theta}=(\pmb{K},P)$, and $\pmb{\Theta}=(\pmb{\theta},\pmb{\alpha})$. By independence, we see that the probability of edge assignment  between a class-$i$ node $v_x$ and a class-$j$ node $v_y$ in $\mathbb{H}(n;\pmb{\mu},\pmb{\Theta})$ is given by
\begin{equation} \nonumber
\mathbb{P}[v_x \sim v_y \mid t_x=i,t_y=j]=\alpha_{ij} p_{ij}
\end{equation}
Similar to (\ref{eq:osy_mean_edge_prob_in_RKG}), we denote the mean edge probability for a class-$i$ node in $\mathbb{H}(n;\pmb{\mu},\pmb{\Theta})$ as $\Lambda_i$. It is clear that
\begin{align} 
\Lambda_i = \sum_{j=1}^r \mu_j \alpha_{ij} p_{ij}, \quad i=1,\ldots, r.
\label{eq:osy_mean_edge_prob_in_system}
\end{align}
We denote the minimum mean edge probability in $\mathbb{H}(n;\pmb{\mu},\pmb{\Theta})$ as $\Lambda_m$, i.e., 
\begin{equation*}
m:=\arg \min_i \Lambda_i.
\label{eq:min_mean_degree}
\end{equation*}
We also let
\begin{align}
&d:=\arg \max_j \alpha_{mj}, \label{eq:HetER0}\\
&s:=\arg \max_j \alpha_{mj} p_{mj}. \label{eq:HetER_s}
\end{align}

Throughout, we assume that the number of classes $r$ is fixed and does not scale with $n$, and so are the probabilities $\mu_1, \ldots,\mu_r$. All of the remaining parameters are assumed to be scaled with $n$.

\section{Main Results and Discussion}
We refer to a mapping $K_1,\ldots,K_r,P: \mathbb{N}_0 \rightarrow \mathbb{N}_0^{r+1}$ as a \textit{scaling} (for the inhomogeneous random key graph) if\begin{equation}
1 \leq K_{1,n} \leq K_{2,n} \leq \ldots \leq K_{r,n} \leq P_n/2
\label{scaling_condition_K}
\end{equation}
hold  for all $n=2,3,\ldots$. Similarly any mapping $\pmb{\alpha}=\{ \alpha_{ij} \}: \mathbb{N}_0 \rightarrow (0,1)^{r \times r}$ defines a scaling for the inhomogeneous ER graphs. A mapping $\pmb{\Theta} : \mathbb{N}_0 \rightarrow \mathbb{N}_0^{r+1} \times (0,1)^{r \times r}$ defines a scaling for the intersection graph $\mathbb{H}(n;\pmb{\mu},\pmb{\Theta})$ given that condition (\ref{scaling_condition_K}) holds. We remark that under (\ref{scaling_condition_K}), the edge probabilities $p_{ij}$ will be given by
(\ref{eq:osy_edge_prob_type_ij}).

\subsection{Results}

We present a zero-one law for the absence of isolated nodes in 
$\mathbb{H}(n;\pmb{\mu},\pmb{\Theta}_n)$.
\begin{theorem}
\label{theorem:isolated_nodes}
Consider a probability distribution $\pmb{\mu}=\{\mu_1,\mu_2,\ldots,\mu_r\}$ with $\mu_i >0$ for $i=1,\ldots,r$, a scaling $K_1,\ldots,K_r,P: \mathbb{N}_0 \rightarrow \mathbb{N}_0^{r+1}$, and a scaling $\pmb{\alpha}=\{\alpha_{ij}\}: \mathbb{N}_0 \rightarrow (0,1)^{r \times r}$ such that
\begin{equation}
\Lambda_m(n) \sim c \frac{\log n}{n}
\label{scaling_condition_KG}
\end{equation}
holds for some $c>0$.

i) If
\begin{equation*}
\lim_{n \to \infty} \alpha_{md}(n) \log n=0,
\end{equation*}
or 
\begin{align*}
&\lim_{n \to \infty} \alpha_{md}(n) \log n=\alpha^* \in (0,\infty],\\
&\lim_{n \to \infty} \alpha_{mm}(n) \log n=\alpha^{**} \in (0,\infty].
\end{align*}
Then, we have
\begin{equation} \nonumber
\lim_{n\to\infty} \mathbb{P} \left[ \begin{split} \mathbb{H}(n;\pmb{\mu},\pmb{\Theta}_n) \text{ has} \\ \text{ no isolated nodes} \end{split} \right]= 0 \quad \text{ if } c<1
\end{equation}

ii) We have
\begin{equation} \nonumber
\lim_{n\to\infty} \mathbb{P} \left[ \begin{split} \mathbb{H}(n;\pmb{\mu},\pmb{\Theta}_n) \text{ has} \\ \text{ no isolated nodes} \end{split} \right]= 1 \quad \text{ if } c>1
\end{equation}
\end{theorem}
%
%
%
The scaling condition (\ref{scaling_condition_KG}) will often be used in the form
\begin{equation} \label{scaling_condition_KG_v2}
\Lambda_m(n)=c_n \frac{\log n}{n}, \ n=2,3,\ldots
\end{equation}
with $\lim_{n\to\infty} c_n=c>0$. 

Theorem~\ref{theorem:isolated_nodes} states that $\mathbb{H}(n;\pmb{\mu},\pmb{\Theta}_n)$ has no isolated node whp if the minimum mean degree, i.e., $n \Lambda_m$, is scaled as $(1+\epsilon) \log n$ for some $\epsilon > 0$. On the other hand, if this minimum mean degree scales as $(1-\epsilon) \log n$ for some $\epsilon > 0$, then whp $\mathbb{H}(n;\pmb{\mu},\pmb{\Theta}_n)$ has a class-$m$ node that is isolated, and hence not connected. We remark that  $\alpha^{**} \leq \alpha^*$ since $\alpha_{mm}(n) \leq \alpha_{md}(n)$ for $n=1,2,\ldots$.

The zero-one law established here for the absence of isolated nodes in $\mathbb{H}(n;\pmb{\mu},\pmb{\Theta}_n)$ shall be regarded as a crucial
first step towards establishing the connectivity result.
In fact, Theorem~\ref{theorem:isolated_nodes} 
already implies the zero-law for connectivity, i.e., that
\[
\lim_{n \to \infty} \mathbb{P} \left[ \begin{split} & \mathbb{H}(n;\pmb{\mu},\pmb{\Theta}_n) \\ & \text{is connected} \end{split} \right]= 0   \quad \text{ if } c<1.
\]
This is because a graph can not be connected if it contains an isolated node. Also, 
for several classes of random graphs it is known that the conditions that ensure connectivity coincide with those ensuring absence of isolated nodes; e.g., 
random key graphs \cite{yagan2012zero}, (homogeneous) ER graphs \cite{Bollobas}, and random geometric key graphs \cite{PenroseBook}. 
This prompts us to introduce the following conjecture.

\begin{conjecture}
\label{conj:1}
{\sl
Consider a probability distribution $\pmb{\mu}=(\mu_1,\mu_2,\ldots,\mu_r)$ with $\mu_i >0$ for $i=1,\ldots,r$, a scaling $K_1,\ldots,K_r,P: \mathbb{N}_0 \rightarrow \mathbb{N}_0^{r+1}$, and a scaling $\pmb{\alpha}=\left(\alpha_{ij} \right): \mathbb{N}_0 \rightarrow (0,1)^{r \times r}$ such that (\ref{scaling_condition_KG}) holds.
With either

i) 
\begin{equation*}
\lim_{n \to \infty} \alpha_{md}(n) \log n=0,
\end{equation*}
or

ii) 
\begin{align*}
&\lim_{n \to \infty} \alpha_{md}(n) \log n=\alpha^* \in (0,\infty],\\
&\lim_{n \to \infty} \alpha_{mm}(n) \log n=\alpha^{**} \in (0,\infty],
\end{align*}
 and possibly under some additional conditions, we have
\begin{equation} \nonumber
\lim_{n \to \infty} \mathbb{P} \left[ \begin{split} & \mathbb{H}(n;\pmb{\mu},\pmb{\Theta}_n) \\ & \text{is connected} \end{split} \right]=
\left \{
\begin{array}{cl}
0     &  \text{ if } c<1   \\
1   &    \text{ if } c>1
\end{array}
 \right.
\end{equation}
}
\end{conjecture}

\subsection{Comparison with related work}
Our main result extends the results established by Eletreby and Ya\u{g}an in \cite{Rashad/Inhomo} for the inhomogeneous random key graph intersecting the (homogeneous) ER graph. There, zero-one laws for the property that the graph has no isolated nodes and the property that the graph is connected were established. It is clear that our work generalizes the model given in \cite{Rashad/Inhomo} by considering the \textit{inhomogeneous} ER graph, enabling the analysis of networks with heterogeneous radio capabilities. Indeed, when $\alpha_{ij}(n)=\alpha(n)$ for $i,j=1,\ldots,r$ and each $n=1,2,\ldots$, our result recovers the absence of isolated nodes result given in \cite{Rashad/Inhomo}.

In \cite{Yagan/Inhomogeneous}, zero-one laws for the property that the graph has no isolated nodes and the property that the graph is connected were established for the inhomogeneous random key graph $\mathbb{K}(n,\pmb{\mu},\pmb{K},P)$ under the full visibility assumption. It is clear that, although a crucial first step in the study of heterogeneous key predistribution schemes, the full visibility assumption is not likely to hold in most practical settings. 
In fact, by setting  
$\alpha_{ij}(n)=1$ for $i,j=1,\ldots,r$ and each $n=1,2,\ldots$ (i.e., by assuming that all wireless channels are {\em on}), our absence of isolated nodes result reduces to that given in
\cite{Yagan/Inhomogeneous}. 

Finally, Ya\u{g}an in \cite{Yagan/EG_intersecting_ER} considered the homogeneous random key graph (where all nodes receive $K_n$ keys), intersecting the homogeneous ER graph \cite{ER}.
Our work  generalizes  \cite{Yagan/EG_intersecting_ER} by considering the intersection of the \textit{inhomogeneous} ER graph with a more general random graph model that accounts for the cases where nodes can be assigned different number of keys; i.e., with the \textit{inhomogeneous} random key graph. In fact, with $r=1$, i.e., when $\alpha$ is a scalar and all nodes belong to the same class and thus receive the same number of keys, our
absence of isolated node result recovers the result given in \cite{Yagan/EG_intersecting_ER}.

\subsection{Significance of the results}
\subsubsection{Network Reliability Problem}
The problem studied in this paper is closely connected to the popular {\em network reliability problem} \cite[Section 7.5]{Bollobas}, described as follows:
Starting with a fixed, deterministic graph $\mathcal{H}$, obtain $\mathbb{I}(\mathcal{H};p)$ by deleting each edge of $\mathcal{H}$ independently with probability $1-p$.
Network reliability problem is often translated to finding the probability that $\mathbb{I}(\mathcal{H};p)$ is connected as a function of $p$. For arbitrary graphs, $\mathcal{H}$, this problem is shown 
\cite{Valiant,ProvanBall}
to be $\#P$-complete, meaning that no polynomial algorithm exists for its solution, unless $P=NP$. Our result given above constitutes a crucial first step towards the asymptotic solution of the network reliability problem for inhomogeneous random key graphs when edges are deleted with \textit{different} probabilities. Put differently, we consider a generic network reliability problem; wherein, different links fail with different probabilities which paves the way for many interesting problems. Although asymptotic in nature, our result can still provide useful insights about the reliability properties of random key graphs with number of vertices $n$ being on the order of thousands.

\subsubsection{Common-Interest Friendship Networks} 
We demonstrate an application of our result in the context of a common-interest friendship network, denoted by $G_c$. A common interest relationship between two friends manifests from their selection of common objects from a pool of available objects. Clearly, this is modeled by the inhomogeneous random key graph; wherein, the inhomogeneity captures the fact that different people have different number of interests. The friendship network is modeled by an inhomogeneous ER graph, meaning that any two individuals are connected with a probability that is based on their corresponding \textit{classes} independently from other individuals. The class of an individual could represent her job title, current city, academic degree, etc. As a result, $G_c$ becomes the intersection of the inhomogeneous random key graph with the inhomogeneous ER graph. Our results on its absence of isolated nodes constitutes the first step in revealing the conditions under which global information diffusion can take place in the common-interest network. In particular, when $G_c$ is connected, global information diffusion is possible.

\section{Numerical Results}
\label{sec:numerical}
We now present numerical results and simulations to check the validity of Theorem~\ref{theorem:isolated_nodes} in the finite node regime. In all experiments, we fix the number of nodes at $n = 500$ and the size of the key pool at $P = 10^4$. For better visualization, we use the curve fitting tool of MATLAB.

\begin{figure}[t]
\centerline{\includegraphics[scale=0.45]{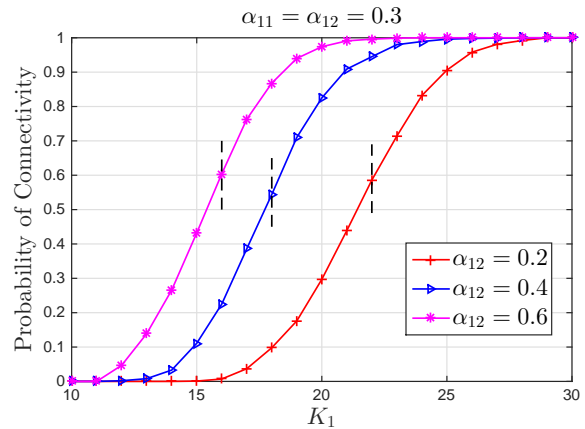}}
\caption{Empirical probability that $\mathbb{H}(n;\pmb{\mu},\pmb{\Theta})$ is connected as a function of $\pmb{K}$ for $\alpha_{12} = 0.2$, $\alpha_{12} = 0.4$, and $\alpha_{12} = 0.6$ with $n = 500$ and $P = 10^4$; in each case, $\alpha_{11}=\alpha_{22}=0.3$. The empirical probability value is obtained by averaging over $400$ experiments. Vertical dashed lines stand for the critical threshold of connectivity asserted by Conjecture~\ref{conj:1}.}
\label{fig:1}
\end{figure}

In Figure~\ref{fig:1}, we set the channel matrix to
\begin{equation*}
\pmb{\alpha}=
  \begin{bmatrix}
    0.3 & \alpha_{12} \\
    \alpha_{12} & 0.3
  \end{bmatrix}
\end{equation*}
and consider the channel parameters $\alpha_{12} = 0.2$, $\alpha_{12} = 0.4$, and $\alpha_{12} = 0.6$, while varying the parameter $K_1$ (i.e., the smallest key ring size) from $10$ to $35$. The number of classes is fixed to $2$, with $\pmb{\mu}=\{0.5,0.5\}$. For each value of $K_1$, we set $K_2=K_1+5$. For each parameter pair $(\pmb{K}, \pmb{\alpha})$, we generate $400$ independent samples of the graph $\mathbb{H}(n;\pmb{\mu},\pmb{\Theta})$ and count the number of times (out of a possible $400$) that the obtained graphs i) have no isolated nodes and ii) are connected. Dividing the counts by $400$, we obtain the (empirical) probabilities for the events of interest.  In all cases considered here, we observe that $\mathbb{H}(n;\pmb{\mu},\pmb{\Theta})$ is connected whenever it has no isolated nodes yielding the same empirical probability for both events. This confirms the asymptotic equivalence of the connectivity and absence of isolated nodes properties in $\mathbb{H}(n;\pmb{\mu},\pmb{\Theta}_n)$ as we give by Conjecture~\ref{conj:1}.

For each value of $\alpha_{12}$, we show the critical threshold of connectivity \lq\lq predicted" by Conjecture~\ref{conj:1} by a vertical dashed line. More specifically, the vertical dashed lines stand for the minimum integer value of $K_1$ that satisfies
\begin{equation}
\Lambda_m(n)=\sum_{j=1}^2 \mu_j \alpha_{mj} \left( 1- \frac{\binom{P-K_j}{K_m}}{\binom{P}{K_m}} \right) >\frac{\log n}{n}.
\label{eq:numerical_critical}
\end{equation}
We see from Figure~\ref{fig:1} that the probability of connectivity transitions from zero to one within relatively small variations of $K_1$. Moreover, the critical values of $K_1$ obtained by (\ref{eq:numerical_critical}) lie within this transition interval. We finally note that for each parameter pair $(\pmb{K},\pmb{\alpha})$ in Fig~\ref{fig:1}, we have $\Lambda_m=\Lambda_1$.

\begin{figure}[t]
\centerline{\includegraphics[scale=0.45]{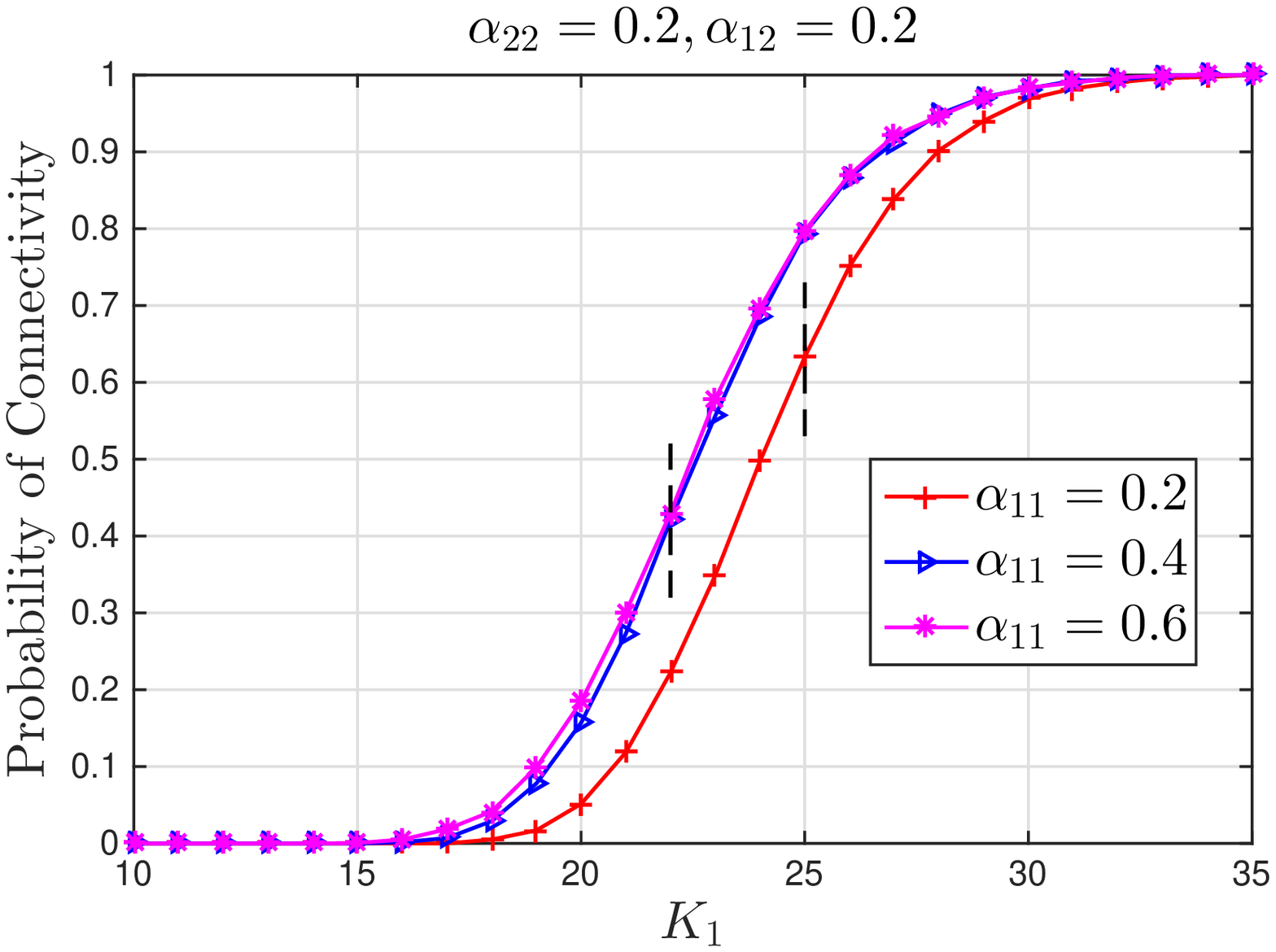}}
\caption{Empirical probability that $\mathbb{H}(n;\pmb{\mu},\pmb{\Theta})$ is connected as a function of $\pmb{K}$ for $\alpha_{11} = 0.2$, $\alpha_{11} = 0.4$, and $\alpha_{11} = 0.6$ with $n = 500$ and $P = 10^4$; in each case, $\alpha_{12}=\alpha_{22}=0.2$. The empirical probability value is obtained by averaging over $400$ experiments. Vertical dashed lines stand for the critical threshold of connectivity asserted by Conjecture~\ref{conj:1}.}
\label{fig:2}
\end{figure}

Next, we set the channel matrix to
\begin{equation*}
\pmb{\alpha}=
  \begin{bmatrix}
    \alpha_{11} & 0.2 \\
    0.2 & 0.2
  \end{bmatrix}
\end{equation*}
in Figure~\ref{fig:2}, and consider the channel parameters $\alpha_{11} = 0.2$, $\alpha_{11} = 0.4$, and $\alpha_{11} = 0.6$, while varying the parameter $K_1$ (i.e., the smallest key ring size) from $10$ to $35$. The number of classes is fixed to $2$, with $\pmb{\mu}=\{0.5,0.5\}$. For each value of $K_1$, we set $K_2=K_1+5$. Using the same procedure that produced Figure~\ref{fig:1}, we obtain the empirical probability that $\mathbb{H}(n;\pmb{\mu},\pmb{\Theta})$ is connected versus $K_1$. As before, the critical threshold of connectivity asserted by Conjecture~\ref{conj:1} is shown by a vertical dashed line in each curve. One interesting observation of Figure~\ref{fig:2} is how the behavior of the probability of connectivity changes with $\alpha_{11}$. In fact, when $\alpha_{11}=0.2$, we have $\Lambda_m=\Lambda_1$, while for $\alpha_{11} \geq 0.4$, we have $\Lambda_m=\Lambda_2$. Consequently, the value of $\alpha_{11}$ (which only appears in the calculations of $\Lambda_1$) becomes irrelevant to the scaling condition given by (\ref{eq:numerical_critical}). We notice from Fig~\ref{fig:2}, that for $\alpha_{11} \geq 0.4$, fixed $\alpha_{12}$, and fixed $\alpha_{22}$, we have the same critical value of $K_1$ and quite similar behavior of the probability of connectivity. 

\begin{figure}[t]
\centerline{\includegraphics[scale=0.45]{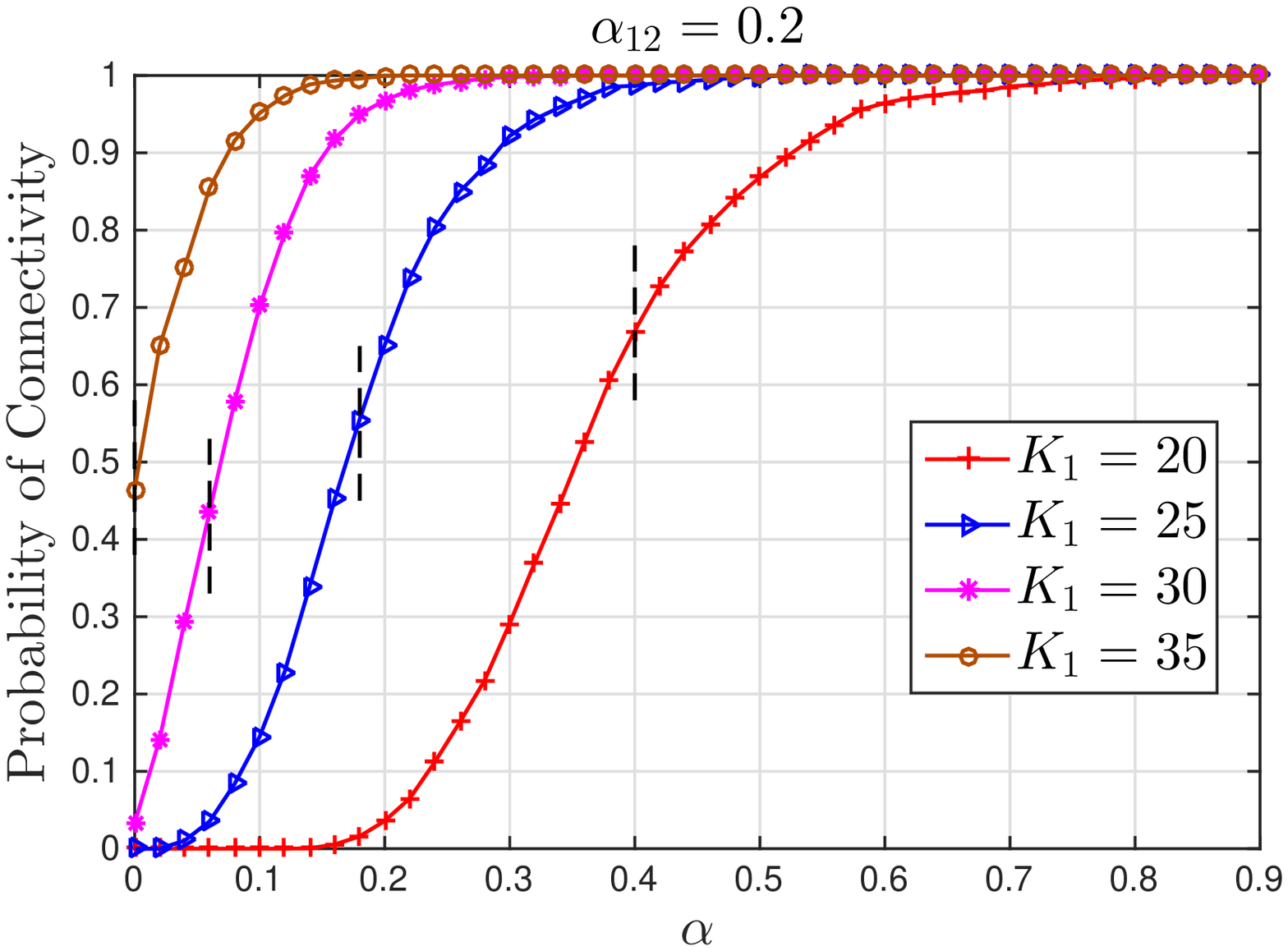}}
\caption{Empirical probability that $\mathbb{H}(n;\pmb{\mu},\pmb{\Theta})$ is connected as a function of $\alpha$ for $K_1 = 20$, $K_1 = 25$, $K_1 = 30$, and $K_1=35$, with $n = 500$ and $P = 10^4$; in each case, $\alpha_{12}=0.2$ . The empirical probability value is obtained by averaging over $400$ experiments. Vertical dashed lines stand for the critical threshold of connectivity asserted by Conjecture~\ref{conj:1}.}
\label{fig:3}
\end{figure}

\begin{figure}[t]
\centerline{\includegraphics[scale=0.45]{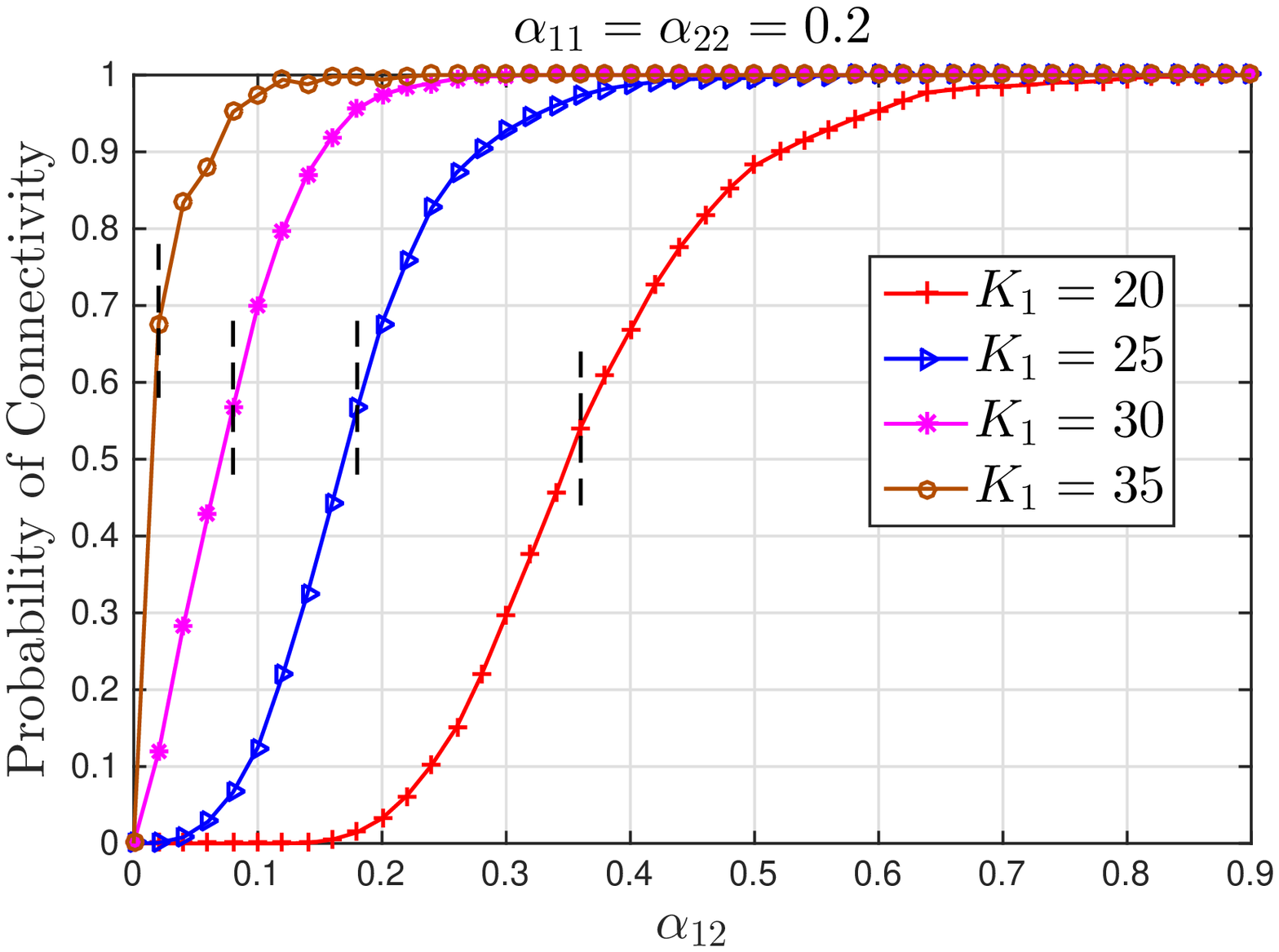}}
\caption{Empirical probability that $\mathbb{H}(n;\pmb{\mu},\pmb{\Theta})$ is connected as a function of $\alpha_{12}$ for $K_1 = 20$, $K_1 = 25$, $K_1 = 30$, and $K_1=35$, with $n = 500$ and $P = 10^4$; in each case, $\alpha_{11}=\alpha_{22}=0.2$ . The empirical probability value is obtained by averaging over $400$ experiments. Vertical dashed lines stand for the critical threshold of connectivity asserted by Conjecture~\ref{conj:1}.}
\label{fig:4}
\end{figure}

Finally, we set the channel matrix to
\begin{equation*}
\pmb{\alpha}=
  \begin{bmatrix}
    \alpha & 0.2 \\
    0.2 & \alpha
  \end{bmatrix}
\end{equation*}
and consider four different minimum key ring sizes, $K_1 = 20$, $K_1 = 25$, $K_1 = 30$, and $K_1=50$ while varying the parameter $\alpha$ from $0$ to $1$. The number of classes is fixed to $2$ with $\pmb{\mu}=\{0.5,0.5\}$ and we set $K_2=K_1+5$ for each value of $K_1$. Using the same procedure that produced Figure~\ref{fig:1}, we obtain the empirical probability that $\mathbb{H}(n;\pmb{\mu},\pmb{\Theta})$ is connected versus $\alpha$. As before, the critical threshold of connectivity asserted by Conjecture~\ref{conj:1} is shown by a vertical dashed line in each curve. One interesting observation from Figure~\ref{fig:3} is that $\mathbb{H}(n;\pmb{\mu},\pmb{\Theta})$ could possibly be connected with $\alpha_{12}>0$ even when $\alpha=0$. In particular, the resultant graph becomes a \textit{bipartite} graph; namely, class-$1$ nodes are adjacent only to class-$2$ nodes and class-$2$ nodes are adjacent only to class-$1$ nodes. Such a behavior confirms the importance of $\alpha_{12}$ over $\alpha_{11}$ and $\alpha_{22}$. This is also captured in Figure~\ref{fig:4}; wherein, the probability of connectivity is indeed $0$ when $\alpha_{12}=0$.

\section{Proof of Theorem~\ref{theorem:isolated_nodes}}
\label{sec:proof_isolated}
\subsection{Preliminaries}
Several technical results are collected here for convenience.
The first result follows easily from the scaling condition (\ref{scaling_condition_K}). 
\begin{prop} [{\cite[Proposition~4.1]{Yagan/Inhomogeneous}}]
For any scaling $K_1,K_2,\ldots,K_r,P:\mathbb{N}_0 \rightarrow \mathbb{N}_0^{r+1}$, we have
\begin{equation}
\lambda_1(n) \leq \lambda_2(n) \leq \ldots \leq \lambda_r(n)
\label{eq:isolated_ordering_of_lambda}
\end{equation}
for each $n=2,3,\ldots$.
\end{prop}

{\prop[{\cite[Proposition~4.4]{Yagan/Inhomogeneous}}]
For any set of positive integers $K_1,\ldots,K_r,P$ and any scalar $a \geq 1$, we have
\begin{equation}
\frac{\binom {P-\left \lceil{aK_i}\right \rceil }{K_j}}{\binom P{K_j}} \leq
\left(\frac{\binom {P-K_i}{K_j}}{\binom P{K_j}}\right)^a, \quad i,j=1,\ldots,r
\label{eq:isolated:combinatorial_bound}
\end{equation}
}

Other useful bound that will be used throughout is
\begin{align}
& (1 \pm x) \leq e^{\pm x}, \quad x \in (0,1)
\label{eq:isolated:exp_bound}
\end{align}

Finally, we find it useful to write
\begin{equation}
\log (1-x)=-x-\Psi(x)
\label{eq:isolated_log_decomp}
\end{equation}
where  $\Psi(x)=\int_{0}^{x} \frac{t}{1-t} \ \text{dt}$.
From L'H\^{o}pital's Rule, we have
\begin{equation}
\lim_{x\to 0}  \frac{\Psi(x)}{x^2}=\frac{-x-\log (1-x)}{x^2}=\frac{1}{2}.
\label{eq:isolated_hopital}
\end{equation}

\subsection{Establishing the one-law}
The proof of Theorem~\ref{theorem:isolated_nodes} relies on the method of first and second moments applied to the number of isolated nodes in $\mathbb{H}(n;\pmb{\mu},\pmb{\Theta}_n)$. Let $I_n(\pmb{\mu},\pmb{\Theta}_n)$ denote the total number of isolated nodes in $\mathbb{H}(n;\pmb{\mu},\pmb{\Theta}_n)$, namely,
\begin{equation}
I_n(\pmb{\mu},\pmb{\Theta}_n)=\sum_{\ell=1}^n \pmb{1}[v_\ell \text{ is isolated in }\mathbb{H}(n;\pmb{\mu},\pmb{\Theta}_n)]
\label{eq:isolated_In}
\end{equation}
The method of first moment \cite[Eqn. (3.10), p.
55]{JansonLuczakRucinski} gives
\begin{equation} \nonumber
1-\mathbb{E}[I_n(\pmb{\mu},\pmb{\Theta}_n)]\leq \mathbb{P}[I_n(\pmb{\mu},\pmb{\Theta}_n)=0] 
\end{equation}

It is clear that in order to establish the one-law, namely that $ \lim_{n \to \infty} \mathbb{P}\left[  I_n(\pmb{\mu},\pmb{\Theta_n})=0\right]=1$, we need to show that
\begin{equation*}
\lim_{n \to \infty} \mathbb{E}[I_n(\pmb{\mu},\pmb{\Theta}_n)]=0.
\end{equation*}

Recalling (\ref{eq:isolated_In}), we have
\begin{align}
&\mathbb{E}\left[I_n(\pmb{\mu},\pmb{\Theta}_n)\right] \nonumber \\
&=n \sum_{i=1}^r \mu_i \mathbb{P}\left[v_1 \text{ is isolated in }\mathbb{H}(n;\pmb{\mu},\pmb{\Theta}_n) \given[\big] t_1=i\right]\nonumber \\
&=n \sum_{i=1}^r \mu_i \mathbb{P}\left[\cap_{j=2}^n [v_j \nsim v_1] \mid v_1 \text{ is class i}\right]\nonumber \\
&=n \sum_{i=1}^r \mu_i \left(\mathbb{P}\left[v_2 \nsim v_1 \mid v_1 \text{ is class i}\right]\right)^{n-1} \label{eq:isolated_independence}
\end{align}
where (\ref{eq:isolated_independence}) follows by the independence of the rvs $\{v_j \nsim v_1\}_{j=1}^n$ given $\Sigma_1$. By conditioning on the class of $v_2$, we find
\begin{align}
\mathbb{P}[v_2 \nsim v_1 \given[\big] t_1=i]
&=\sum_{j=1}^r \mu_j \mathbb{P}[v_2 \nsim v_1 \given[\big] t_1=i,t_2=j]\nonumber \\
&=\sum_{j=1}^r \mu_j (1-\alpha_{ij} p_{ij})=1-\Lambda_i(n).
\label{eq:isolated_OneLaw_first_step}
\end{align}
Using (\ref{eq:isolated_OneLaw_first_step}) in (\ref{eq:isolated_independence}), and recalling (\ref{eq:min_mean_degree}),  (\ref{eq:isolated:exp_bound}) we obtain
\begin{align*}
\mathbb{E}[I_n(\pmb{\mu},\pmb{\Theta}_n)] &= n \sum_{i=1}^r \mu_i \left(1-\Lambda_i(n)\right)^{n-1}\nonumber \\
&\leq n \left(1-\Lambda_m(n)\right)^{n-1}\nonumber \\
&= n \left(1-c_n \frac{\log n}{n}\right)^{n-1}\nonumber \\
&\leq  e^{\log n \left(1-c_n  \frac{n-1}{n}\right)}
\end{align*}
Taking the limit as $n$ goes to infinity, we immediately get
\begin{equation*}
\lim_{n \to \infty} \mathbb{E}[I_n(\pmb{\mu},\pmb{\Theta}_n)]=0.
\end{equation*}
since $\lim_{n \to \infty} (1-c_n  \frac{n-1}{n})=1-c < 0$ under the enforced assumptions (with $c>1$) and the one-law is established.

\subsection{Establishing the zero-law}
Our approach in establishing the zero-law relies on the method of second moment applied to a variable that counts the number of nodes that are class-$m$ and isolated. Clearly if we can show that whp there exists at least one class-$m$ node that is isolated under the enforced assumptions (with $c<1$) then the zero-law would immediately follow.

Let $Y_n(\pmb{\mu},\pmb{\Theta}_n)$ denote the number of nodes that are class-$m$ and isolated in $\mathbb{H}(n;\pmb{\mu},\pmb{\Theta}_n)$, and let
\begin{equation} \nonumber
x_{n,i}(\pmb{\mu},\pmb{\Theta}_n)=\pmb{1}[t_i=m \cap v_i \text{ is isolated in }\mathbb{H}(n;\pmb{\mu},\pmb{\Theta}_n)],
\end{equation}
then we have $Y_n(\pmb{\mu},\pmb{\Theta}_n)=\sum_{i=1}^n x_{n,i}(\pmb{\mu},\pmb{\Theta}_n)$. By applying the method of second moments
\cite[Remark 3.1, p. 55]{JansonLuczakRucinski}  on $Y_n(\pmb{\mu},\pmb{\Theta}_n)$, we get
\begin{equation}
\mathbb{P}[Y_n(\pmb{\mu},\pmb{\Theta}_n)=0] \leq 1-\frac{\mathbb{E}[Y_n(\pmb{\mu},\pmb{\Theta}_n)]^2}{\mathbb{E}[Y_n(\pmb{\mu},\pmb{\Theta}_n)^2]}  
\label{eq:isolated_ZeroLaw_bound}
\end{equation}
where
\begin{equation}
\mathbb{E}[Y_n(\pmb{\mu},\pmb{\Theta}_n)]=n \mathbb{E}[x_{n,1}(\pmb{\mu},\pmb{\Theta}_n)]
\label{eq:isolated_ZeroLaw_first_part}
\end{equation}
and
\begin{align}
\begin{split}
\mathbb{E}[Y_n(\pmb{\mu},\pmb{\Theta}_n)^2]=&n \mathbb{E}[x_{n,1}(\pmb{\mu},\pmb{\Theta}_n)]\\
&+n(n-1)\mathbb{E}[x_{n,1}(\pmb{\mu},\pmb{\Theta}_n) x_{n,2}(\pmb{\mu},\pmb{\Theta}_n)]
\end{split}
\label{eq:isolated_ZeroLaw_second_part}
\end{align}
by exchangeability and the binary nature of the rvs $\{x_{n,i}(\pmb{\mu},\pmb{\Theta}_n) \}_{i=1}^n$.
Using (\ref{eq:isolated_ZeroLaw_first_part}) and  (\ref{eq:isolated_ZeroLaw_second_part}), we get
\begin{equation} \nonumber
\begin{split}
\frac{\mathbb{E}[Y_n(\pmb{\mu},\pmb{\Theta}_n)^2]}{\mathbb{E}[Y_n(\pmb{\mu},\pmb{\Theta}_n)]^2}  =& \frac{1}{n \mathbb{E}[x_{n,1}(\pmb{\mu},\pmb{\Theta}_n)]} \\
&+ {\frac{n-1}{n} \frac{\mathbb{E}[x_{n,1}(\pmb{\mu},\pmb{\Theta}_n) x_{n,2}(\pmb{\mu},\pmb{\Theta}_n)]}{\mathbb{E}[x_{n,1}(\pmb{\mu},\pmb{\Theta}_n)]^2}}
\end{split}
\end{equation}

In order to establish the zero-law, we need to show that
\begin{equation}
\lim_{n \to \infty} n \mathbb{E}[x_{n,1}(\pmb{\mu},\pmb{\Theta}_n)]= \infty,
\label{eq:isolated_ZeroLaw_first_condition}
\end{equation}
and
\begin{equation}
\limsup_{n \to \infty} \left(\frac{\mathbb{E}[x_{n,1}(\pmb{\mu},\pmb{\Theta}_n) x_{n,2}(\pmb{\mu},\pmb{\Theta}_n)]}{\mathbb{E}[x_{n,1}(\pmb{\mu},\pmb{\Theta}_n)]^2}\right) \leq 1.
\label{eq:isolated_ZeroLaw_second_condition}
\end{equation}

We establish (\ref{eq:isolated_ZeroLaw_first_condition}) and (\ref{eq:isolated_ZeroLaw_second_condition}) in the following propositions.
{\prop
\label{prop:prop1_osy}
Consider a scaling $K_1,\ldots,K_r,P:\mathbb{N}_0 \rightarrow \mathbb{N}_0^{r+1}$ and a scaling $\pmb{\alpha}=\{\alpha_{ij}\}:=\mathbb{N}_0 \rightarrow (0,1)^{r \times r}$ such that (\ref{scaling_condition_KG}) holds with $\lim_{n \to \infty} c_n=c>0$. Then, we have
\begin{equation*}
\lim_{n \to \infty} n \mathbb{E}[x_{n,1}(\pmb{\mu},\pmb{\Theta}_n)]= \infty, \quad \text{if } c<1
\end{equation*}
}
\begin{proof}
We have
\begin{align}
&n \mathbb{E}\left[x_{n,1}(\pmb{\mu},\pmb{\Theta}_n)\right] \nonumber \\
&=n \mathbb{E}\left[\pmb{1}[t_1=m \cap v_1 \text{ is isolated in }\mathbb{H}(n;\pmb{\mu},\pmb{\Theta}_n)]\right]\nonumber \\
&=n \mu_m \mathbb{P}\left[\cap_{j=2}^n [v_j \nsim v_1] \given[\big] t_1=m\right]\nonumber \\
&=n \mu_m \mathbb{P}\left[v_2 \nsim v_1\given[\big] t_1=m\right]^{n-1}\nonumber \\
&=n \mu_m \left(\sum_{j=1}^r \mu_j (1-\alpha_{mj} p_{mj})\right)^{n-1} \nonumber \\
&=n \mu_m \left(1-\Lambda_m(n)\right)^{n-1} = \mu_m e^{\beta_n} 
\label{eq:isolated_ZeroLaw_simp1}
\end{align}
where 
\begin{equation*}
\beta_n=\log n+(n-1)\log (1-\Lambda_m(n)).\\
\end{equation*}
Recalling (\ref{eq:isolated_log_decomp}), we get
\begin{align}
\beta_n&=\log n-(n-1)\left(\Lambda_m(n)+\Psi(\Lambda_m(n))\right)\nonumber \\
&=\log n \left(1-c_n \frac{n-1}{n}\right) \nonumber \\
& \quad -(n-1) \left(c_n \frac{\log n}{n} \right)^2 \frac{\Psi \left(c_n \frac{\log n}{n}\right)}{\left(c_n \frac{\log n}{n}\right)^2}
 \label{eq:isolated_ZeroLaw_simp2}
\end{align}

Recalling (\ref{eq:isolated_hopital}), we have
\begin{equation}
\lim_{n \to \infty} \frac{\Psi \left(c_n \frac{\log n}{n}\right)}{\left(c_n \frac{\log n}{n}\right)^2} = \frac{1}{2}
\label{eq:isolated_ZeroLaw_simp3}
\end{equation}
since $c_n \frac{\log n}{n}=o(1)$. Thus, $\beta_n=\log n \left(1-c_n \frac{n-1}{n}\right)-o(1)$.
Using (\ref{eq:isolated_ZeroLaw_simp1}), (\ref{eq:isolated_ZeroLaw_simp2}), (\ref{eq:isolated_ZeroLaw_simp3}), and letting $n$ go to infinity, we get
\begin{equation*}
\lim_{n \to \infty} n \mathbb{E}[x_{n,1}(\pmb{\mu},\pmb{\Theta}_n)]= \infty
\end{equation*}
whenever $\lim_{n \to \infty} c_n=c < 1$.
\end{proof}

{\prop
Consider a scaling $K_1,\ldots,K_r,P:\mathbb{N}_0 \rightarrow \mathbb{N}_0^{r+1}$ and a scaling $\pmb{\alpha}=\{\alpha_{ij}\}:=\mathbb{N}_0 \rightarrow (0,1)^{r \times r}$ such that (\ref{scaling_condition_KG}) holds with $\lim_{n \to \infty} c_n=c>0$. Then, we have (\ref{eq:isolated_ZeroLaw_second_condition}) if $c<1$.
\label{prop:new_osy}
}
We omit the proof of Proposition~\ref{prop:new_osy} from this conference version. All details can be found in \cite{Eletreby_AllertonFullVersion}. Collectively, Proposition~\ref{prop:prop1_osy} and Proposition~\ref{prop:new_osy} establish (\ref{eq:isolated_ZeroLaw_first_condition}) and (\ref{eq:isolated_ZeroLaw_second_condition}) which in turn establish the zero-law.

\section*{Acknowledgment}
This work has been supported in part by National Science Foundation through grants CCF-1617934 and CCF-1422165
and in part by the start-up funds from the Department of Electrical and Computer Engineering at Carnegie Mellon University. 

\ifCLASSOPTIONcaptionsoff
  \newpage
\fi

\bibliographystyle{IEEEtran}
\bibliography{IEEEabrv,Allerton_Final}

\end{document}